\documentclass[12pt]{article}
\usepackage{epsf}
\title{ {\bf CP-Violation in  $b\rightarrow s\ell^{+}\ell^{-}$ transition Beyond the
Standard Model}}
\author{\vspace{1cm}\\
 V. Bashiry \thanks {e-mail:
bashiry@newton.physics.metu.edu.tr}\\ ENGINEERING FACULTY, CYPRUS
INTERNATIONAL UNIVERSITY,\\ Via Mersin 10 , Turkey}

\begin{document}
\setlength{\baselineskip}{24pt} \maketitle
\setlength{\baselineskip}{7mm}
\begin{abstract}
In this study the CP-asymmetry in the $b\rightarrow
s\ell^{+}\ell^{-}$ transition was investigated in minimal extension
of the Standard Model where $C_{9}^{eff}$ receives an extra weak
phase due to the new physics effects.  We observed that CP-Violation
asymmetry can be measurable in the framework of scenario mentioned
above.
\end{abstract}
\thispagestyle{empty}
\newpage
\setcounter{page}{1}
\section{Introduction}
 Rare B meson decay induced by the flavor changing neutral current (FCNC)$b\rightarrow s(d)$
 transition provides potentially the most sensitive and precise
 test for the standard model (SM) in the flavor sector at loop
 level since such transitions are forbidden in SM at tree level.
 At the same time, these decays are very sensitive to the new
 physics beyond the SM.\\
Experimental investigation of these decays will provide a more
precise determination of the elements of the
Cabbibo-Kobayashi-Maskawa (CKM) matrix such as ,$V_{tb},\,
V_{ts},V_{td}$ and $V_{ub}$ . Moreover , they can provide better
insight
into understanding the origin of CP-violation. \\
 Rare semileptonic decays $b\rightarrow
 s(d)\ell^{+}\ell^{-}$ are more informative for this aim, since
these decays are relatively clean compared to pure hadronic decays.
Note that the semileptonic $b\rightarrow q \ell^+ \ell^-$ transition
has been extensively studied in numerous works
\cite{R5704}--\cite{bashiry} in the framework of the SM and its
various  extensions. The matrix elements of the $b\rightarrow
s\ell^{+}\ell^{-}$ transition contain terms describing the virtual
effects induced by the $\overline{t}t$, $\overline{c}c$ and
$\overline{u}u$ loops, which are proportional to
$|V_{tb}V^{*}_{ts}|$, $|V_{cb}V^{*}_{cs}|$ and $|V_{ub}V^{*}_{us}|$,
respectively. Using the unitarity condition of the CKM matrix and
recalling that one can neglect $|V_{ub}V^{*}_{us}|$ in comparison to
$|V_{tb}V^{*}_{ts}|$ and $|V_{cb}V^{*}_{cs}|$, it is obvious that
the matrix element for the $b\rightarrow s\ell^{+}\ell^{-}$
transition involves only one independent CKM matrix element, namely
$|V_{tb}V^{*}_{ts}|$, so the CP-violation in this channel is
strongly suppressed in the SM. However, the possibility of
CP-violation as the result of new physics effects in $b\rightarrow
s$ transition has been studied in
supersymmetry\cite{Kruger:2000zg}-\cite{Kruger:2000ff}, and
 in SM with fourth generation\cite{Arhrib:2002md}. In another study, this has been studied
 with the addition of CP odd phases to Wilson coefficients\cite{Aliev:2005pw}.\\
Situation for $b\rightarrow d\ell^{+}\ell^{-}$ is totaly different
from $b\rightarrow s\ell^{+}\ell^{-}$ transition. In this case, all
CKM matrix elements $|V_{td}V^{*}_{tb}|$, $|V_{cd}V^{*}_{cb}|$ and
$|V_{ud}V^{*}_{ub}|$ are in the same order and for this reason the
matrix element of $b\rightarrow d\ell^{+}\ell^{-}$ transition
contains two different amplitudes with two different CKM elements
and therefore it is expected to have a large CP
violation.\\
In order to get CP violation not only do we have to have two
different amplitudes but also these amplitudes must contain pieces
which transform under CP transformation in different ways, i.e.
these amplitudes must contain weak and strong phases. It is clear
that the weak phase changes its sign under CP
transformation but strong phase doesn't.\\
Let's briefly recall the situation in $b\rightarrow
s\ell^{+}\ell^{-}$ transition in standard model. In the SM, the
Wilson coefficients $C_7$ and $C_{10}$ are real , while $C_9^{eff}$
contains strong and weak phases. The $C_9^{eff}$ is usually
parameterized in the the following form:

\begin{equation}\label{c9}
C^{eff}_{9}=\xi_{1}+\lambda_{u}\xi_{2}
\end{equation}
where
\begin{equation}\label{lau}
\lambda_{u}= \frac{|V_{ub}V^{*}_{us}|}{|V_{tb}V^{*}_{ts}|}
\end{equation}
As we have noted, this quantity is very small, therefore, it
 is usually neglected in calculations and for this reason CP
violation in
this channel is strongly suppressed.\\
As we mentioned above, the $b\rightarrow s(d)\ell^{+}\ell^{-}$
transition is a promising candidate looking for new physics beyond
the SM. New physics effects can appear in rare decays when the
Wilson coefficients take values different from their SM counterpart
or new operator structures in effective
Hamiltonian which are absent in the SM. \\
In the present work, we focus on CP asymmetry in $ b\rightarrow
s\ell^{+}\ell^{-}$ transition. This asymmetry is very small in SM;
therefore, any deviation of this asymmetry from SM prediction
clearly indicates existence of new physics. We should note that the
first measurement of the $ b\rightarrow s\ell^{+}\ell^{-}$ decay
reported by BELLE \cite{R5720} is:
\begin{equation}\label{brexp}
{\cal{B}}(B \rightarrow X_s\ell^+\ell^-)=(6.1\pm
1.4^{+1.4}_{-1.1})\times10^{-6}.
\end{equation}
 An experiment recently done by BELLE Collaboration\cite{ishik}
 examined the  measurement of forward--backward asymmetry and
determination of Wilson coefficients in $B\rightarrow
K^*\ell^+\ell^-$ decay. In future, a similar measurement will be
done for CP violating asymmetry.
 Although all Wilson
coefficients($C_7^{eff},\,C_9^{eff}$ and $C_{10}^{eff}$) can take CP
odd phases, we will discuss the possibility elsewhere. We will here
discuss the situation where $C_9^{eff}$ will get a new weak phase.
We call this minimal extension of Standard Model since, with this
extension, CP asymmetry can appear in $b\rightarrow s$ transition .
As we consider the minimal extension of the SM, we assume that the
strong phase is the same as the SM case, because it appears in
imaginary part of polarization operator.\\
 This paper is organized as follows: In Section
2 , we present the theoretical framework for the decay width and
CP-violation asymmetry. Section 3 encompasses our numerical results
and an estimation of the feasibility of measuring the CP violation
 and conclusion.
\section{Theoretical Framework}
In this section, we present the theoretical expressions for the
decay widths and CP-violation asymmetry. As we mentioned above, we
restrict ourselves by considering minimal extension of the SM, more
precisely, we extend only $C_9^{eff}$, since in SM only this
coefficient has weak and strong phases, i.e. :
\begin{equation}\label{c9new}
C^{new}_{9}=\xi_{1}+(\lambda_{u}+\lambda_{new})\xi_{2}
\end{equation}
where $\lambda_{u}$ is given by Eq.(\ref{lau}) and $\lambda_{new}$
is parameterized as:
\begin{equation}\label{lanew}
    \lambda_{new}=|\lambda_{new}|\exp(i\varphi)
\end{equation}
The explicit expressions of functions $\xi_1$ and $\xi_2$ are
respectively\cite{burev}--\cite{misiakE}:
\begin{eqnarray}\label{xi1}
\xi_1 & = & 4.128 \, +\, 0.138
\omega(\hat{s})\,+\,g(\hat{m}_{c},\hat{s}) (3 C_1 + C_2 + 3 C_3 +
C_4 + 3 C_5 + C_6)\nonumber\\&-& \frac{1}{2}g(\hat{m}_{d},\hat{s})
(C_3 + C_4) - \frac{1}{2}
   g(\hat{m}_{b},\hat{s})(4 C_3 + 4 C_4 + 3C_5 + C_6) \\
  & + &\frac{2}{9} (3 C_3 + C_4 + 3C_5 + C_6)\nonumber \\
   \xi_2 & = & [ g(\hat{m}_{c},\hat{s})- g(\hat{m}_{u},\hat{s})](3
C_1 + C_2)\label{xi2}
\end{eqnarray}
Where $\hat{m}_q=\,m_q/m_b$ and $\hat{s}=\,\frac{q^2}{m_b^2} $\\
 As we know, the rare decays are one of the promising classes of decays for new physics beyond
the SM.  \\
 The QCD corrected effective Hamiltonian describing $b\rightarrow
s\ell^{+}\ell^{-}$ transitions leads to the matrix element:
\begin{eqnarray}
 M =\frac{G_{F}\alpha
V_{tb}V^{*}_{ts}}{\sqrt{2}\pi}[&C^{new}_{9}(\overline{s}\gamma_{\mu}P_{L}b)\overline{\ell}\gamma_{\mu}\ell
+C_{10}(\overline{s}\gamma_{\mu}P_{L}b)\overline{\ell}\gamma_{\mu}\gamma^{5}\ell
\nonumber\\&-2\,C_{7}^{eff}\overline{s}i\sigma_{\mu\nu}\frac{q^{\nu}}{q^{2}}
(m_{b}P_{R}+m_{s}P_{L})b\overline{\ell}\gamma_{\mu}\ell\,]
 , \,\,\, \label{amplitude}
\end{eqnarray}
 where  $q$ denotes the four
momentum of the lepton pair.
 Neglecting the terms of
$O(m_q^2/m_W^2)$, $q = u, d, c$, the analytic expressions for all
Wilson coefficients, except $C_{9}^{new}$,  can be found in
\cite{misiak, burev}. The values of $C_{7}^{eff}$ and $C_{10}$ in
leading logarithmic approximation are:
\begin{equation} \label{wilsonc7c10}
C_{7}^{eff} = -0.315,\quad  C_{10} =  -4.642
\end{equation}
The function $g(\hat{m}_{q}, \hat{s})$ represents the corrections to
the four-quark operators $O_1-O_6$ \cite{misiakE} and is defined as:
\begin{eqnarray}
g(\hat{m}_{q}, \hat{s}) &=& -\frac{8}{9} \ln(\hat{m}_{q}) +
\frac{8}{27} + \frac{4}{9}\ y_{q}
          - \frac{2}{9} (2 + y_{q}) \sqrt{|1-y_{q}|}\
\bigg\{\Theta(1-y_{q}) \times
\nonumber\\
&&\left[ \ln\left(\frac{1+\sqrt{1-y_{q}}}{1-\sqrt{1-y_{q}}}\right)
- i \pi \right]
 + \Theta(y_{q} - 1)\ 2 \arctan{\frac{1}{\sqrt{y_{q}-1}}}
\bigg\}~, \label{gmq}
\end{eqnarray}
Even though we neglect long-distance resonance effects in this
paper, a more complete analysis of the above decay has to take into
account the long-distance contributions, which have their origin in
real intermediate $c\bar{c}$ family, in addition to the
short-distance contribution. In the case of  the $J/\psi$ family,
this is usually accomplished by introducing a Breit-Wigner
distribution for the resonances through the replacement
\cite{longdist}
 \begin{eqnarray}\label{long-dist}
  g(\hat{m}_{c}, \hat{s}) \longrightarrow
g(\hat{m}_{c}, \hat{s})-\frac{3\pi}{\alpha^2}\sum\limits_{V= J/\psi,
\psi', \dots} \frac{\hat{m}_V Br(V\to l^+l^-)\hat{\Gamma}_{\mathrm
{total}}^V} {\hat{s}-\hat{m}^2_V + i \hat{m}_V \hat{\Gamma}_{\mathrm
{total}}^V}\ ,
\end{eqnarray}
Using the expression of matrix element in equation (\ref{amplitude})
and neglecting the s-quark mass ($m_s$)~\cite{aali}--\cite{falk}, we
obtain the expression for the differential decay rate
as~\cite{babu};
\begin{eqnarray}
\Gamma_0\,=\frac{d \Gamma}{d \hat{s}} = \frac{G_F m_b^5}{192
\pi^3} \frac{\alpha^2}{4 \pi^2} |V_{tb} V_{ts}^*|^2 (1 -
\hat{s})^2 \sqrt{1 - \frac{4 \hat{m}_{\ell}^2}{\hat{s}}}
\bigtriangleup \label{difdecaywidth}
\end{eqnarray}
 with
\begin{eqnarray}
\bigtriangleup &=& 4 \frac{(2 + \hat{s})}{\hat{s}} \left(1 +
\frac{2 \hat{m}_{\ell}^2}{\hat{s}}\right) |C_7^{eff}|^2 + (1 + 2
\hat{s}) \left(1 + \frac{2 \hat{m}_{\ell}^2}{\hat{s}}\right)
|C_9^{new}|^2 \nonumber \\
&& + (1 - 8 \hat{m}_{\ell}^2 + 2 \hat{s} + \frac{2
\hat{m}_{\ell}^2}{\hat{s}}) |C_{10}|^2  + 12 (1 + \frac{2
\hat{m}_{\ell}^2}{\hat{s}}) Re(C_9^{new *} C_7^{eff})  .
\label{delta}
\end{eqnarray}
In the unpolarized case, the CP-Violating asymmetry rate can be
defined by
\begin{eqnarray}
A_{cp}(\hat{s})=\frac{\Gamma_0\,-\,\overline{\Gamma}_0}{\Gamma_0\,+\,\overline{\Gamma}_0}
\label{acps}
\end{eqnarray}
 where
\begin{eqnarray}
\Gamma_0\equiv\frac{d\Gamma}{d\hat{s}}\equiv\frac{d\Gamma(b\rightarrow
s \ell^{+}\ell^{-})}{d\hat{s}},\quad
\overline{\Gamma}_0\equiv\frac{d\overline{\Gamma}}{d\hat{s}}\equiv\frac{d\overline{\Gamma}(\overline{b}\rightarrow
\overline{s} \ell^{+}\ell^{-})}{d\hat{s}} \label{gama0}
\end{eqnarray}

 The explicit expression for the unpolarized particle decay rate
$\Gamma_0$ has been given in (\ref{difdecaywidth}). Obviously, it
can be written as a product of a real-valued function $r(\hat{s})$
times the function $\Delta(\hat{s})$, given in (\ref{delta});
$\Gamma_0(\hat{s}) = r(\hat{s}) \ \Delta(\hat{s})$. Taking the
approach in \cite{kruger}, we write the matrix elements for the
decay and the anti-particle decay as
\begin{eqnarray}
M = A + \lambda_{new} B, \quad \overline{M}= A + \lambda_{new}^{
*}B
\end{eqnarray}
where the CP-violating parameter $\lambda_{new}$, entering the
Wilson coupling ${C^{new}_{9}}$, has been defined in Eq.
(\ref{c9new}). Consequently, the rate for the anti-particle decay
is, then, given by
\begin{equation}
\bar \Gamma_0 = {\Gamma_0}_{|\lambda_{new} \to \lambda_{new}^*} =
r(\hat{s}) \bar \Delta(\hat{s})~;
~\bar\Delta=\Delta_{|{\lambda_{new} \to \lambda_{new}^*}}~.
\label{apr}
\end{equation}
Using (\ref{difdecaywidth}) and (\ref{apr}), the CP violating
asymmetry is evaluated to be \cite{kruger}
\begin{equation}\label{asymcomp}
A_{CP}(\hat{s}) = \frac{ -2\,  |\lambda_{new}|\sin(\varphi)
\Sigma}{\Delta + 2\, |\lambda_{new}|\cos(\varphi)\Sigma}.
\end{equation}
Furthermore, we can do the same approximation done for $b\rightarrow
d \ell^+ \ell^-$ \cite{bashiry,babu,kruger}, if the
$\lambda_{new}\sim 1$. So, we can ignore the term proportional to
the $\Sigma$ in the dominator of Eq.(\ref{asymcomp}):
\begin{equation}\label{asymcomp1}
A_{CP}(\hat{s}) \approx  -2\,  |\lambda_{new}|\sin(\varphi)
\frac{\Sigma}{\Delta }.
\end{equation}
 Where $\Delta$ is defined by Eq.(\ref{delta}) and
$\Sigma$ is as follows:
\begin{eqnarray}\label{sigma}
\Sigma&=& Im [\xi_1^* \,\xi_2]f_+(\hat{s}) + Im (C_7^{eff}
\xi_2^{})
f_1(\hat{s})\nonumber\\
f_+(\hat{s})&=&(1 + 2 \hat{s}) \left(1 + \frac{2\hat{m}_{\ell}^2}{\hat{s}}\right)\nonumber\\
f_1(\hat{s})&=& 12 (1 + \frac{2 \hat{m}_{\ell}^2}{\hat{s}})
\end{eqnarray}
\section{Numerical analysis}
 In this section, we examine the dependence of CP-violating
asymmetry  on $\lambda_{new}$ .\\ From the Eq.(\ref{acps}) it
follows that CP violating asymmetry depends on both  magnitude and
phase of $\lambda_{new}$ . In order to have an idea about magnitude
of $\lambda_{new}$, we assume that the normalized branching ratio
can departure from SM result Eq.(\ref{brexp}) by about 10 percent in
the presence of the new parameters, i.e,
\begin{equation}\label{restbr}
    -0.1\leq\frac{B_{r}^{new}-B_{r}^{SM}}{B_{r}^{SM}}\leq 0.1
\end{equation}
Note that, the same approach  used in \cite{bashirychin}. Solving
this equation on $|\lambda_{new}|$, we obtain the upper limit for
$|\lambda_{new}|\leq1.16 $, so the allowed region for
$|\lambda_{new}|$ is:
\begin{equation}\label{restla}
0.0\leq|\lambda_{new}|\leq1.16
\end{equation}
We, here, see that, under the condition  mentioned above,
$|\lambda_{new}|$ has to be zero for SM case.     \\
 The values of input parameters which we
use in our numerical analysis
 are: $ m_{s}=0.15$ GeV, $ m_{b}=4.8$ GeV,
 $ m_{\mu}=105.7$ MeV,  $ C_{7}=-0.314$,  $C_{10}=-4.642 $, $ \alpha=\frac{1}{129}$
 \cite{PDG}. Moreover, we use Wolfenstein parametrization \cite{wolf} for
 CKM matrix. The current values of the Wolfenstein
 parameters are $A=0.83$ and $\lambda=0.221$ \cite{robert}.
Besides, we use the following simple expression of $\xi_1$ and
$\xi_2$ in the NLO approximation \cite{babu}
\begin{equation}
\xi_1 \simeq 4.128 + 0.138\ \omega(\hat{s}) + 0.36\ g(\hat{m}_c,
\hat{s}),~~ \xi_2 \simeq 0.36\ [g(\hat{m}_c, \hat{s}) -
g(\hat{m}_u, \hat{s})]~. \label{approx}
\end{equation}
And the explicit expression of the $\omega(\hat{s})$ is:
\begin{eqnarray}\label{omegas}
\omega(\hat{s}) & =&  -\frac{2}{9} \pi^2
-\frac{4}{3}Li_2(\hat{s})-\frac{2}{3} \ln (\hat{s})
\ln(1-\hat{s})-\frac{5+4\hat{s}}{3(1+2\hat{s})} \ln(1-\hat{s}) \nonumber\\
&-& \frac{2
\hat{s}(1+\hat{s})(1-2\hat{s})}{3(1-\hat{s})^2(1+2\hat{s})} \ln
(\hat{s}) + \frac{5+9 \hat{s}-6
\hat{s}^2}{3(1-\hat{s})(1+2\hat{s})}~,
\end{eqnarray}
In order to eliminate the $\hat{s}$ dependence instead of CP
asymmetry in differential decay width, we study CP asymmetry in
total decay width by doing numerical integration over $\hat{s}$ in
Eq. (\ref{asymcomp1}).
\begin{equation}\label{ACP}
A_{CP} = -2\,
|\lambda_{new}|\sin(\varphi)\frac{\int^{(1-m_s^2/m_b^2)}_{(4m^2_{\mu})/m_b^2}
\Sigma\,d\hat{s} }{\int^{(1-m_s^2/m_b^2)}_{(4m^2_{\mu})/m_b^2}\Delta
d\hat{s}}.
\end{equation}
 In the figure 1, we present the dependence of CP
violating asymmetry on $|\lambda_{new}|$ and $\varphi$ , where
$|\lambda_{new}|$ varies in the region presented by
Eq.{\ref{restla}}. The figure depicts that $A_{CP}$ is sensitive to
the new weak phase and can reach  about $~\%4.5$ percent. For
nominal asymmetry of $\%5$ and branching ratio of $10^{-6}$, a
measurement at $3\sigma$ level requires about $~10^9$ B mesons
\cite{babu,kruger}. In view of clear dilepton signal, such a
measurement is quite feasible at future colliders like LHCb, BTeV,
ATLAS  CMS \cite{Harnew:1999sq} or ILC \cite{Shrihari}. For
instance, it is expected to be produced $10^{12}\,\,\,B\,
\overline{B}$ pairs at LHC. So they will be able to measure
$b\rightarrow s\ell^{+}\ell^{-}$ or exclusive process $B\rightarrow
K(K^*)\ell^{+}\ell^{-}$. Moreover, the existence of this CP
asymmetry in $b\rightarrow s$ transition can be a direct indication
of new physics effects since, in SM, this CP asymmetry is near zero
\cite{bashiry, kruger}. Here a few words about the synergy of LHC
and ILC are in order: It is clear that LHC will reach  higher
energies and can create much more $B\, \overline{B}$ pairs than ILC.
The ILC, on the other hand, can make precision measurements and can
be sensitive to the indirect effects of the new particles which can
contribute to the penguin diagrams of $b\rightarrow s$ transition
even if masses are much higher than the energy of the ILC
\cite{rohini}.

In conclusion, this study presented the  CP-asymmetry in the
$b\rightarrow s\ell^{+}\ell^{-}$ transition in minimal extension of
the Standard Model where $C_{9}^{eff}$ received extra weak phase
$\lambda_{new}$ due to the new physics effects. We imposed $\%10$ of
uncertainty to the SM branching ratio of $b\rightarrow
s\ell^{+}\ell^{-}$ transition and  obtained the bound on new
parameter $\lambda_{new}$. Our predictive model showed that the
CP-Violation asymmetry could reach to the order of $~\%4.5$ which
was not only entirely measurable in experiments, but also indicated
the new physics effects since, in SM, this CP asymmetry is near
zero.
\section{Acknowledgement}
The author thanks  TM Aliev and D. Wyler for their helpful
discussions.

\newpage
{\bf Figure Caption}\\
{\bf Fig. (1)}. The dependence of CP asymmetry $A_{CP}$ on new
parameter $\lambda_{new}$  for the $b\rightarrow s\,\mu^{+} \mu^{-}$
transition. \\
\begin{figure}[1]
\vskip -4.0truein \centering \epsfxsize=5.8in
\leavevmode\epsffile{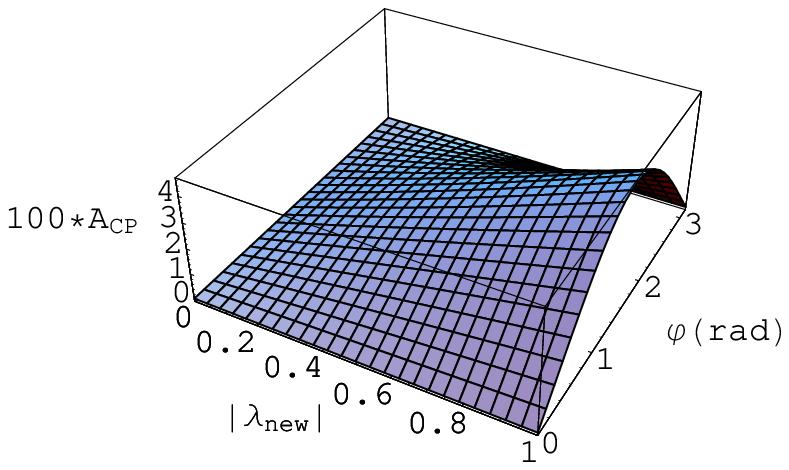} \vskip -0.0truein \caption[]{}
\label{acpfig}
\end{figure}
\end{document}